# Optimization of light yield by injection of the optical filler into the co-extruded hole of plastic scintillation bar with WLS fiber in it

A. Artikov[1], V. Baranov, Yu. Budagov, D. Chokheli[2], Yu. Davydov, V. Glagolev, Yu. Kharzheev[*], V. Kolomoetz, A. Shalugin, A. Simonenko, V. Tereshchenko

*Joint Institute for Nuclear Research, Dzhelepov Laboratory of Nuclear Problems,*

*Joliot-Curie 6, 141980 Dubna, Moscow region, Russian Federation*

e-mail address: Yuri Kharzheev <kharzheev@jinr.ru>

## Abstract

Results of the measurements with cosmic muons for the light yield of 2-meter long extruded scintillation bar (strip) as a function of distance for different options for light collection technique are presented. Scintillation strip cross section geometry was a triangle made on polystyrene plastic scintillator with dopants of 2% PTP and 0.03% POPOP, extruded with 2.6 mm diameter hole and produced at ISMA (Kharkov, Ukraine).

It was shown that the insertion of the optical transparent resin (BC-600 or CKTN-MED(E) ) by special technique into the co-extruded hole with 1.0 mm or 1.2 mm wave-length shifter (WLS) fiber Kuraray Y11 (200) MC in it significantly improves light collection by factor of 1.6…1.9 against of the "dry" case.

---

[1] On leave from NPL, Samarkand State University, Uzbekistan
[2] On leave from IHEP, Tbilisi State University, Georgia
[*] corresponding author

# Content

## 1. Introduction

Detectors based on extruded plastic scintillation bars (strip) are widely used in High Energy Physics (HEP) experiments, particularly, in most of neutrino experiments and will be used in incoming experiments [1]. Usually, strips have a rectangular or triangular shape in cross section of few square centimeters and a length in a few meters: 8 m (MINOS [2]), 7 m (Mu2e [3]), 6 m (OPERA [4]), 3.5 m (MINERvA [5]) and 3 m (T2K [6]). They are co-extruded with groove(s) or hole(s).

Efficiency of detectors based on plastic scintillation counters with wave-length shifter (WLS) fibers is determined by the light yield of the "Scintillation strip – WLS fiber – Photo detector (PD)" system. WLS fiber plays very important role in such systems. It partially absorbs blue light produced by scintillator, re-emits it to the longer wave light (according dopant) and transmits it to PD. Reemitted light lies usually on green light spectra (for Y11) and meets to spectral sensibility of the most PDs: bialkali PMT and SiPM. The light transmission to PD is performed by total internal reflection from the surfaces of the WLS fiber claddings (usually double cladding). Polymethyl methacrylate (PMMA) and fluorinated polymer (FP) are used for inner and outer cladding respectively. The most world famous WLS fibers manufactures are Kuraray (Japan) [7] and BICRON (USA) [8]. Kuraray WLS fibers Y11 (200) MC have emission light spectra maximum 476 nm and attenuation length more than 3.5 m [7]. In general, light readout for bars with WLS fiber less than the 3-meter in length is situated on one side of the bar and other, unread end, is usually polished and then mirrored by depositing a pure Al with epoxy coating. In case of longer scintillator bars – readout is made on the both ends.

WLS fibers are usually inserted into the groove or the hole in the strip. For bars with co-extruded grooves on one of the surfaces (or more), in aim to optimize light contact between bar and WLS fiber, a fiber is often coupled to scintillation at the groove by high transparency optical epoxy cement and value of the refractive index (n) of which is close to the one of the scintillation bar. Such technique (optical coupling) gives the increment of light yield by a factor up to 1.8 [2]. In case of bars with co-extruded hole, WLS fiber usually simply inserted into the hole which means air optical contact inside of the hole between PS and fiber: so called "dry" strip.

We have carried out an investigation on cosmic muons for the optimization of the light collection from 2 m long triangle shape extruded scintillation strip with the co-extruded hole with WLS fiber in. The hole was filled in with some optical resins (optical glue without hardener) by the developed special technique. We have tested 1.0mm and 1.2mm diameter Kuraray Y11 (200) MC WLS fibers. Both ends of fibers were polished and readout end was coupled to PMT EMI9814B with/without optical grease while another end was covered by aluminum Mylar.

1. Design overview

We have tested 2-meter long scintillation strips (Fig. 1) with triangle shape in cross section (with 33 mm as a base and 17 mm in height) made on polystyrene (PS) plastic scintillator with dopants of 2% PTP and 0.03% POPOP extruded at ISMA (Institute for Scintillation Materials, Kharkov, Ukraine) [9]. Strips were covered by $TiO_2$ reflective material and have 2.6 mm diameter hole on its center. Strips with similar shape have been widely used in many experiments (MINERvA [5], T2K [6], D0 [10] et.al.) to arrange large size scintillation counter systems without "dead" zone.

Very high efficiency (need to obtain value of 99,99%) of scintillation counters is required in some HEP experiments, for instance, in Mu2e [3]. High quality for the optical coupling of the WLS fiber to the scintillation strip is one of the perspective ways to attain it. The optical grease for coupling of the WLS fiber end with the PMT window may produce a supplementary increase in light collection.

In our case, WLS fiber was installed into the co-extruded hole of the strip by the insertion it into the hole and, then, the ends of the fiber were fixed by glue at the edges of the strip; the hole was later filled by the optical resin of various types and the optical grease between the end of the WLS fiber and PMT window was accomplished. Finally, light yield collected on PMT was studied on cosmic muons crossing strip at different distances from PMT.

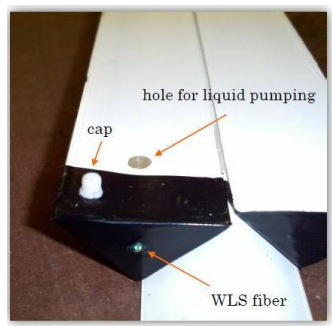

**Fig. 1**. View of 2-meter strip with co-extruded hole and WLS fiber installed in it. Two small holes are used: one for pumping filler (is shown) and another for air expression (not showed). Strips have a triangle shape on cross section with the base of 33 mm, height of 17 mm and the hole on the center with 2.6 mm in diameter.

## 2. Investigation with short strip

Different types of optical transparent fillers for the short length strips (50 cm) were used for first set of tests:

1) Distilled water – low viscosity (n = 1.33 at $20^0$ C);
2) Medical glycerin of 46% aqueous solution (n = 1.39);
3) Ultra-low viscosity "Spectrum-K-59EN" UV glue (n = 1.46) [11]
4) CKTN-MED(E) - low molecular weight and high viscosity synthetic silicone (rubber) (n = 1.606) [12]

The first three fillers have low viscosity (< 20 mPa*s), so inserting them into the strip's hole was easily done by syringe. The forth of them (resin of CKTN(E)) has high viscosity 10…20 Pa*s so we had to develop special technique for filling it into the strip hole (see details in chapter 4 or in [13]).

Results of study of the light yield collection for 50 cm long strips filled by these optical fillers showed an increment in the light yield up to 40-50% against to the "dry" strip case and the best result was obtained with using the rubber (Fig. 2, [13]). Experimental data were fitted by exponential function.

Once the study of light yield for short strips with these fillers in has been performed, we decided to switch focus on study of light yield for 2 m long strip filled by resin of CKTN-MED(E). CKTN-MED is produced in various modifications having the same optical properties but with different viscosity. Two modifications of CKTN-MED with high viscosity are presented in Table 1. We specially chose version E having the most value of viscosity against other modification in order to test possibility of handling with high viscosity filler.

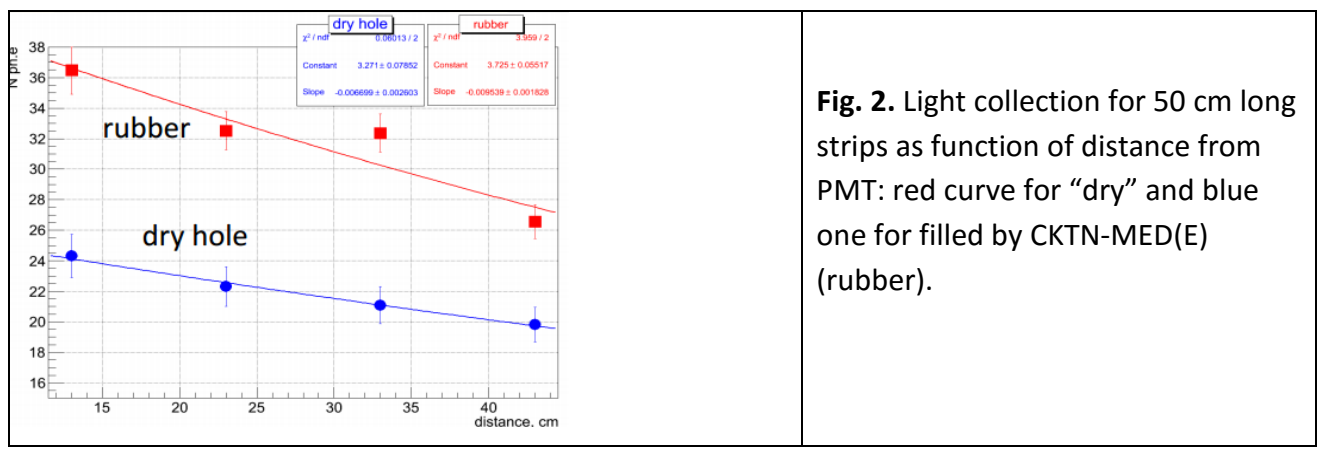

**Fig. 2.** Light collection for 50 cm long strips as function of distance from PMT: red curve for "dry" and blue one for filled by CKTN-MED(E) (rubber).

## 3. High viscosity optical glues as a filler

Different optical glues such as epoxy, silicon and synthetic (based on rubber) have been used for optical coupling. Epoxy glues (for example BC-600 [8]) are often used for gluing WLS fibers on the edge of scintillation bars or into grooves on strips or on tiles in many experiments such as [4], [6], [7], [14], [15] et.al. Epoxy resin Shell EPON 815C was used in experiment MINOS [2] and epoxy Araldite 2020 (corporation Huntsman) is also used elsewhere. Silicone based optical glue SUREL SL-1 [12] (St. Petersburg, Russia) was used for Belle II end-cup KLM detector [16]. However, resin based on epoxy may lose its flexibility under high temperature variations, mechanical influences and by time.

**Table1**. Properties of some optical glue

| Item | Refractive index, n | Transmittance, T, % at $\lambda \sim 500$nm | Viscosity, $\mu$, Pa*s | Cost, USD/kg |
|---|---|---|---|---|
| BC-600 | 1.56 | >98 | 0.8 | ~270 [3] |
| Araldite 2020 | 1.553 | >95 | 0.15 | ~140 |
| CKTN-MED(E) | 1.606 | 92-97 | 10-20 | ~30 [1] |
| CKTN-MED(D) | 1.606 | 92-97 | 6-10 | ~30 [1] |

One of the suggested alternatives to these glues is CKTN-MED(E), produced by SUREL Company [12] mentioned above. It is based on low molecular weight synthetic silicone rubber. It has high flexibility and transparency: light absorption, scattering and transparency for 10 mm thick glue are: 0.4, 0.2 and >95% respectively; it is chemically inert, hydrophobic and its refractive index with n = 1.606 is very close to the one of PS (n = 1.59) for the 500 nm wavelength light. In additional, CKTN-MED is cheaper than BC-600 (Table 1).

Light undergoes reflection while crossing the boundary of 2 mating substances (with refractive indices $n_1$ and $n_2$) and so may be lost. The scale of losses depends on a ratio of its refractive indices, on the light incidence angle, etc. In case of the normal incidence, it may be evaluated by Fresnel equation for the reflection R: $R = [(n_1-n_2) / (n_1+n_2)]^2$. So, optimal light transmission for an optical system including different substances would be achieved if their refractive index matches each other as much as possible.

Optical resins BC-600 and CKTN-MED(E), have been used as filler in hole of PS strip, just well meet these requirements. Moreover, these fillers have a good adhesive with PS. Light transmission through Kuraray Y11(200) MC WLS fiber to PD occurs efficiently by the total internal reflection due to its reflective index of core and inner and outer claddings have 1.59, 1.49, 1.42 respectively.

Inserting a glue into the hole is difficult matter since the ordinary optical glue has a high viscosity. For an example, dynamical viscosity of BC-600 is 0.8 Pa*s and of CKTN-MED(E) is 10…20 Pa*s. This is a reason why WLS fibers are usually inserted in "dry" strip. In such case, it is desirable that co-extruded holes and fiber diameter values to be close to each other as much as possible. For instance, diameters of the hole and WLS fiber were 0.891 mm and 0.835 mm respectively for preshow detectors at D0 experiment [10]; 1.8 mm and 1.5 mm respectively for SciBar detector K2K experiment [17]. However, for strips above 3 m in length, it is difficult to meet this demand: the hole/fiber diameter's ratio may attain about factor 2 and more, and so the filling by the suitable optical filler into the hole can be demanded.

---

[3] January 2016, private communications

We restricted our study by using only the base (resin) of the two-component glues without using hardener because dependence of viscosity on the polymerization time is not well known. It (dependence) should to be clearing up: from our point of view, it is desirable to have an enough time to establish the filling procedure properly before the polymerization.

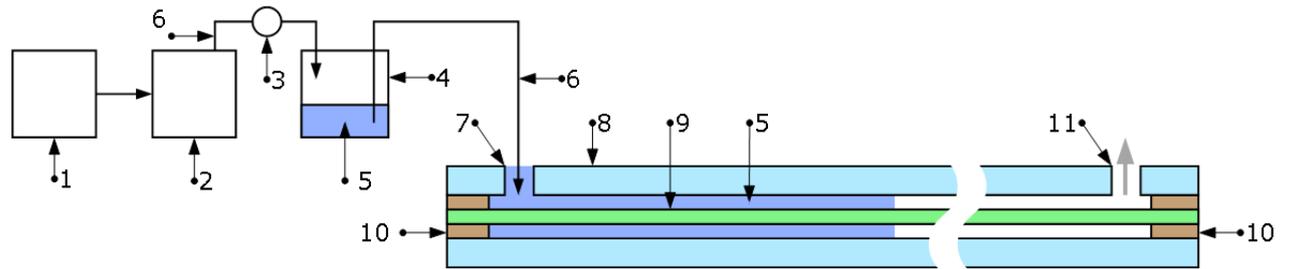

Fig.3. Setup to pump the high viscosity filler into the co-extruded hole of the scintillation bar (no scale): dry type compressor (1); digital Liquid Dispenser SL101N (2); manometer (3); special vessel with filler (4); filler (5); polyvinylchloride tube (6); inlet for filling (7); strip (8); WLS fiber (9); sealing (10); exhaust outlet for extracting air (11).

Special setup was designed and created to pump high viscosity filler into the strip hole (fig. 3). The dry type compressor (1) creates initial pressure and Digital Liquid Dispenser SL101N (2) provides constant value of the pressure on output (about 0.2…0.4 bar over the normal). Pressure is monitored by manometer (3). This constant pressure is provided to the special vessel (4) with the optical filler in it (5) and forces the filler to flow out through the tube (6) and the inlet (7) into the hole of the strip (8). It describes the injection of filler into the strip hole with WLS fiber (9) in it. It takes about 2 hours for 2 m strip under small overpressure of 0.2 bar. Both ends of WLS fibers are glued at the strip edges before filling procedure (10). Air from strip hole is released through exhaust outlet (11).

The reflective coating of the strip sample with 50 cm long was preliminary removed to check visually the filling process. We observed escaping air bubbles inside of co-extruded strip hole while filling. Once the filling process was completed, no bubbles observed along the hole and high adhesive properties of the resin with the fiber and the hole surface were visually demonstrated.

## 4. Investigation with high viscosity fillers

Optical resins CKTN-MED(E) and BC-600 have been chosen as fillers for a long strip. BC-600, well-known and widely used in various detectors for the optical coupling, for instance, to glue the fiber into the grooves on the strips or tiles, was considered as reference to silicon based rubber CKTN-MED (E).

We started the study of the light yield for 2 m long strips with triangle shape filled by resin of CKTN-MED(E) or BC-600 and WLS fiber already installed. The measured light yield from "dry" strips was used as a reference. And tests of 2 m strips, as well as 50 cm long strips, were performed in 2 stages: light yield measurements from "dry state" were done as first and, then, another with same strip, but filled by the resin. Readout was, as described above, performed from one end by coupling of 1.0 mm or 1.2 mm diameter WLS fibers Kuraray Y11(200) MC with PMT EMI 9814B [18]. Both ends of fibers were polished, but the far end of the fiber was covered by Al Mylar besides of a special case when it was covered by black paper.

Transmission of the light from the WLS fiber end to PMT is followed by reflection on their boundary, so various optical coupling compounds usually applied to optimize the light transmission. We have used an optical coupling composition Dow Corning 20-057 (grease) [19] with refractive index 1.48 which is close to that of the PMT window (refractive index of quartz glass is about $n_{quartz}$ = 1.5). Air layer was used as reference as well.

Setup for light yield measurements for the long strips by cosmic muons is shown on Fig. 4. Light was collected by PMT EMI9814B (1) with bialkali photocathode (active diameter 46 mm) and quantum efficiency about 15% at 500 nm (Fig. 4c). Readout strip (3) was placed inside of two light isolated Al tubes (4) covered by black papers inside (not shown). Cosmic rays telescope was comprised of four pairs of plastic scintillation counters (5) with dimensions 20x25x30 mm$^3$ coupled to PMT FEU85 (6) [20]. High voltage supplied to PMT EMI9814B and FEU85 was 2200V and 1000V respectively. Measurements were performed at 8 distances (60, 70, 90, 110, 140, 150, 170, 190 cm) from PMT in 2 stages.

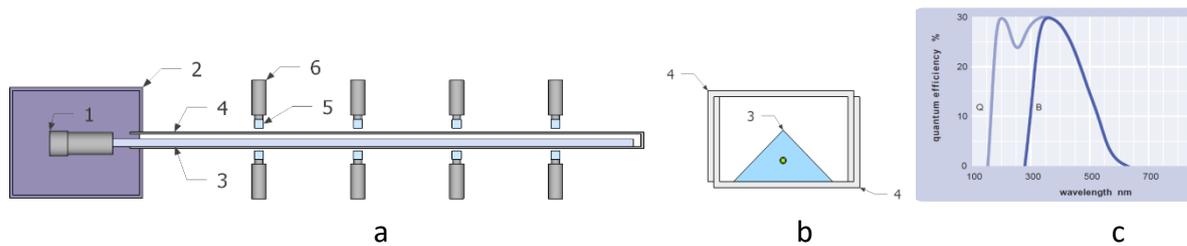

Fig.4. Scheme of experimental setup (a), cross section of light isolated Al tubes with strip inside (b) and spectral sensitivity of PMT EMI9814B (c). PMT EMI9814B (1) in black box (2), strip (3) in light isolated Al tubes (4), 4 pairs of trigger scintillation counters 20x25x30 mm (5) with PMT FEU 85 (6).

Electronic block diagram of the experiment is shown at fig. 5. Analog signal from PMT is measured by a charge-to-digital converter LeCroy ADC 2249W. Signals from cosmic telescope are discriminated by LeCroy 623B; coincidence module LeCroy622 creates output signal to mark position by input register Jorway 65 and runs gate generator LeCroy 222C to produce strobe signal with 100 ns width.

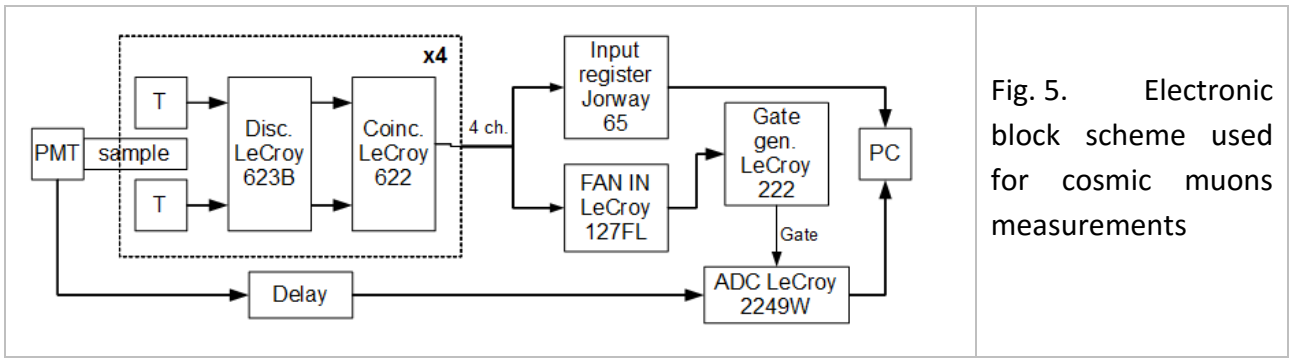

Fig. 5. Electronic block scheme used for cosmic muons measurements

The spectrometric channel was calibrated in absolute units (number of photoelectrons) [21]. Calibration was made by means of a light emission diode (LED) "NICHIA" NSPB310A [22] using light flashes low intensity incident on the PMT photocathode. LED spectrum was used to determine the channel parameters – one photoelectron position and calibration parameter K. A number of photoelectrons were calculated by formula:

$$N_{ph.e.} = (\langle Q \rangle - \langle Q_0 \rangle) K_{att} / K,$$

where: $N_{ph.e}$, $\langle Q \rangle$ and $\langle Q_0 \rangle$ are: a number of photoelectrons (ph. e.), the average spectrum amplitude and pedestal respectively and $K_{att}$ is attenuation coefficient used for increasing dynamical range of ADC scale.

One of the typical calibration spectra obtained by LED is shown (for example) on Fig. 6 where one photoelectron is clearly observed (two photoelectrons position is visible too). LED calibrations have been made before and at the end of every individual run.

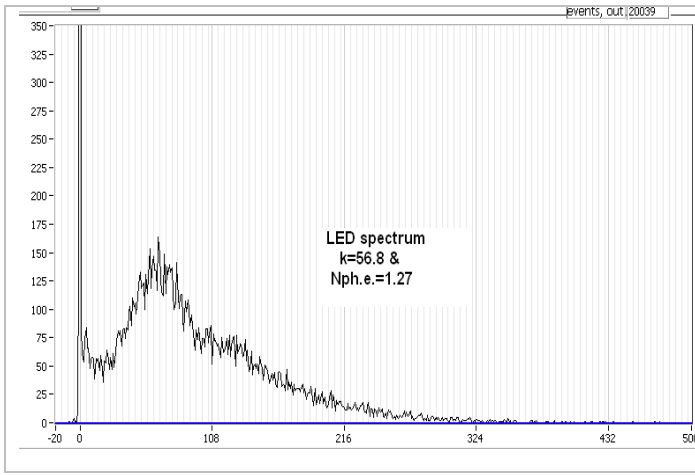

Fig.6. Typical LED spectra

## 5. Results of measurements

Study of the dependence of the light yield for the strip with WLS fiber at the different distances between the trigger counters and PMT, as described above, was carried out and diverse options such as: various fillers and diameters of WLS fibers in the strip hole, contacts between fiber end and PMT with grease or just with air, etc. Results are given on Figures 7-10 and summarized in Tables 2-4. Data presented on Figures 7-10 were fitted by exponential function:

$N_{ph.e} = N_0 * exp(-x/\lambda)$, where λ is technical attenuation length (TAL) of system "strip + WLS fiber" determined as length where intensity of light propagation in such system decrease by a factor e, x is a distance along of strip, $N_{ph.e.}$ is light yield in photoelectrons and $N_0$ is calculated light yield at "$x_0$".

Light yield measurements for the strip with BC-600 as filler were performed for 1.0 mm WLS fiber (Table 2 and Fig. 7) only. One can see that light yield is 1.6 ±0.2 times higher in case of strip filled by BC-600 against of "dry" strip case (see curves 2 and 3 respectively, no optical grease was used to couple WLS fiber and PMT). Optical grease on the touch between PMT and WLS fiber gives supplementary increment in light yield up to 15% (curve 1).

**Table 2**. Light yield (in $N_{ph.e.}$) at different distances for the strip with 1.0 mm WLS fiber and BC-600 as filler in comparison with that of "dry" strip.

| Filler, Φ is diameter of fiber in mm | Distances between trigger counters and PMT, cm | | | | | | | | Grease contact |
|---|---|---|---|---|---|---|---|---|---|
| | 60 | 70 | 90 | 110 | 140 | 150 | 170 | 190 | |
| "dry" strip | 7.9±0.6 | 7.8±0.4 | 7.4±0.5 | 7.0±0.4 | 6.6±0.6 | 6.7±0.4 | 6.6±0.5 | 6.0±0.5 | no |
| BC-600, Φ,1.0 | 13.2±0.8 | 11.9±0.7 | 11.6±0.9 | 11.6±0.8 | 10.7±0.8 | 10.6±0.7 | 10.7±0.8 | 9.4±0.7 | no |
| BC-600, Φ,1.0 | 14.4±0.8 | 13.3±0.9 | 12.3±0.9 | 12.1±0.8 | 11.5±0.8 | 11.3±0.8 | 11.1±0.8 | 9.8±0.7 | yes |

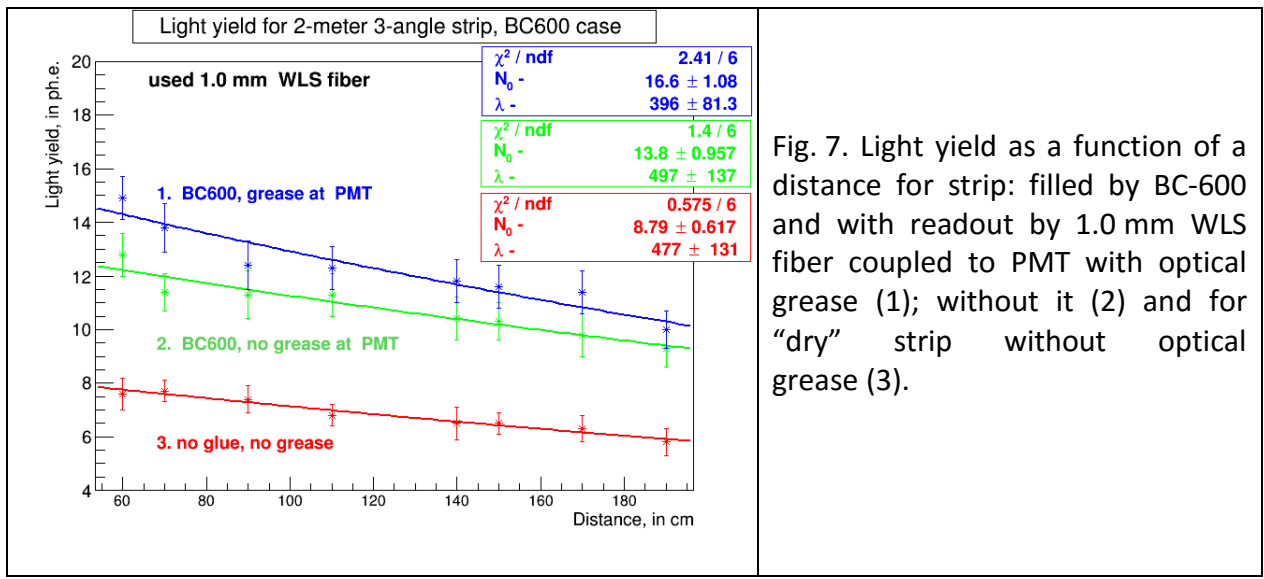

Fig. 7. Light yield as a function of a distance for strip: filled by BC-600 and with readout by 1.0 mm WLS fiber coupled to PMT with optical grease (1); without it (2) and for "dry" strip without optical grease (3).

Light yield measurements for the strip with CKTN-MED(E) and 1.0 mm WLS fiber in its hole (Table 3 and Fig. 8) were done at the same way as that done with BC-600. One can see that light yield is 1.7±0.2 times higher in case strip filled by rubber CKTN-MED(E) against of "dry" strip case (curves 2 and 4 respectively), no optical grease was used to couple WLS fiber and PMT. Using optical grease between PMT and WLS fiber gives supplementary increment for the light yield up to 20% (curves 1).

**Table 3**. Light yield (in $N_{ph.e.}$) at different distances for the strip with 1.0 mm WLS fiber and CKTN-MED(E) as filler in comparison with that of "dry" strip. Data taped in italics script corresponds to measurements performed after half year later than first measurements.

| Filler, Φ is diameter of fiber in mm | Distances between trigger counters and PMT, cm | | | | | | | | Grease contact |
|---|---|---|---|---|---|---|---|---|---|
| | 60 | 70 | 90 | 110 | 140 | 150 | 170 | 190 | |
| "dry" strip | 7.5 ±0.4 | 7.1 ±0.4 | 6.8±0.5 | 6.5 ±0.4 | 6.4 ±0.5 | 6.1 ±0.4 | 6.1 ±0.4 | 5.5 ±0.5 | no |
| CKTN, Φ, 1.0 | 13.0±0.8 | 12.2±0.7 | 11.7±0.7 | 11.3±0.6 | 10.7±0.7 | 10.3±0.6 | 10.1±0.6 | 9.5±0.6 | no |
| CKTN, Φ, 1.0 | 14.8±0.8 | 15.1±0.9 | 13.9±0.8 | 13.0±0.7 | 12.7±0.7 | 12.6±0.7 | 12.2±0.7 | 12.1±0.6 | yes |
| | *14.4±0.8* | *14.7±0.8* | *14.0±0.7* | *12.8±0.7* | *12.4±0.7* | *12.5±0.7* | *12.1±0.7* | *10.1±0.6* | |

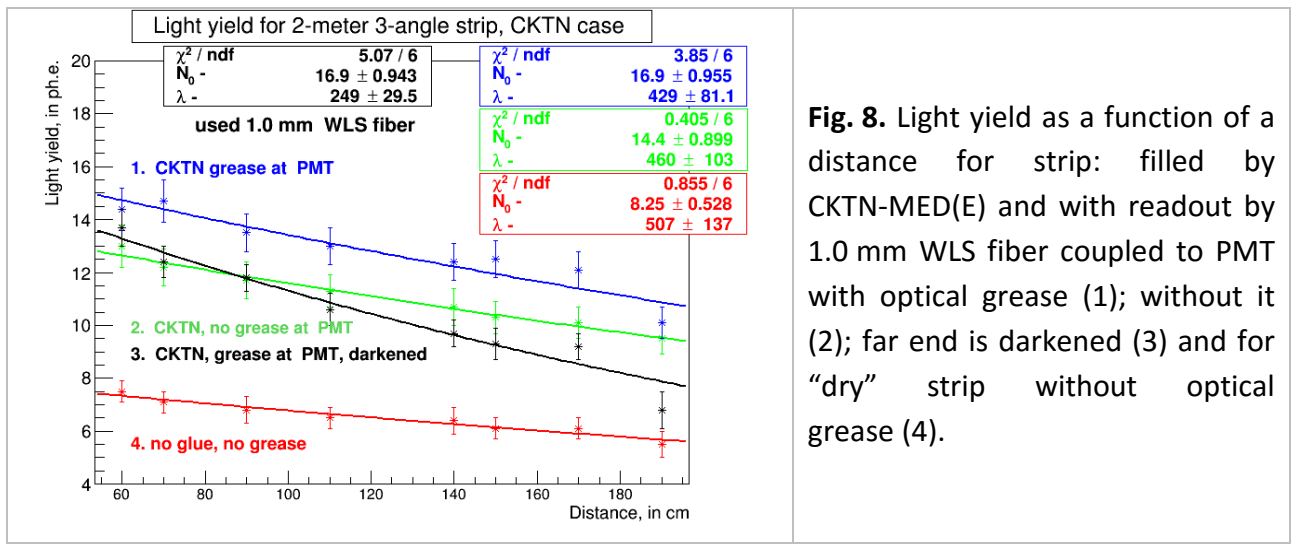

**Fig. 8.** Light yield as a function of a distance for strip: filled by CKTN-MED(E) and with readout by 1.0 mm WLS fiber coupled to PMT with optical grease (1); without it (2); far end is darkened (3) and for "dry" strip without optical grease (4).

Results of light yield for strips with 1.0 mm WLS fiber filled by CKTN-MED(E) or BC-600 has been almost a similar (Tables 2 and 3).

**Table 4.** Light yield (in $N_{ph.e.}$) at different distances for the strip with 1.2 mm WLS fiber and CKTN-MED(E) as filler in comparison with that of "dry" strip.

| Filler, Φ is diameter of fiber, mm | Distances between trigger counters and PMT, cm | | | | | | | | Grease contact |
|---|---|---|---|---|---|---|---|---|---|
| | 60 | 70 | 90 | 110 | 140 | 150 | 170 | 190 | |
| "dry" strip | 13.3±0.8 | 12.3±0.7 | 12.4±0.8 | 11.7±0.7 | 11.2±0.7 | 10.8±0.6 | 10.7±0.7 | 10.0±0.6 | Yes |
| CKTN,Φ,1.2mm | 23.8±1.3 | 23.7±1.3 | 22.7±1.3 | 21.7±1.2 | 21.6±1.2 | 21.0±1.1 | 21.0±1.2 | 16.5±1.0 | Yes |

Light yield measurements of strip with 1.2 mm WLS fiber and CKTN-MED(E) as filler and without it are presented on fig. 9 and Table 4. One can see that the light yield is 1.9±0.2 times higher in case of strip filled by CKTN-MED(E) against of "dry" strip case (curves 1 and 2 respectively); optical grease was used to couple WLS fiber to PMT in both cases.

Similar high increment of 1.8 and 1.6 in the light yield was reached in [2] and [23] respectively. For instance, at MINOS [2], WLS fiber was completely contained and glued inside deep groove on the wide side of rectangular strip and, moreover, reflective aluminum Mylar tape was placed over the groove. With this technique, it was achieved the optical conditions close to that takes in the strip with hole. In case of the study performed in [23], WLS fiber was glued into the groove by the spots in several points along the strip with length of 2.5 m on equal distances (in each 40 cm).

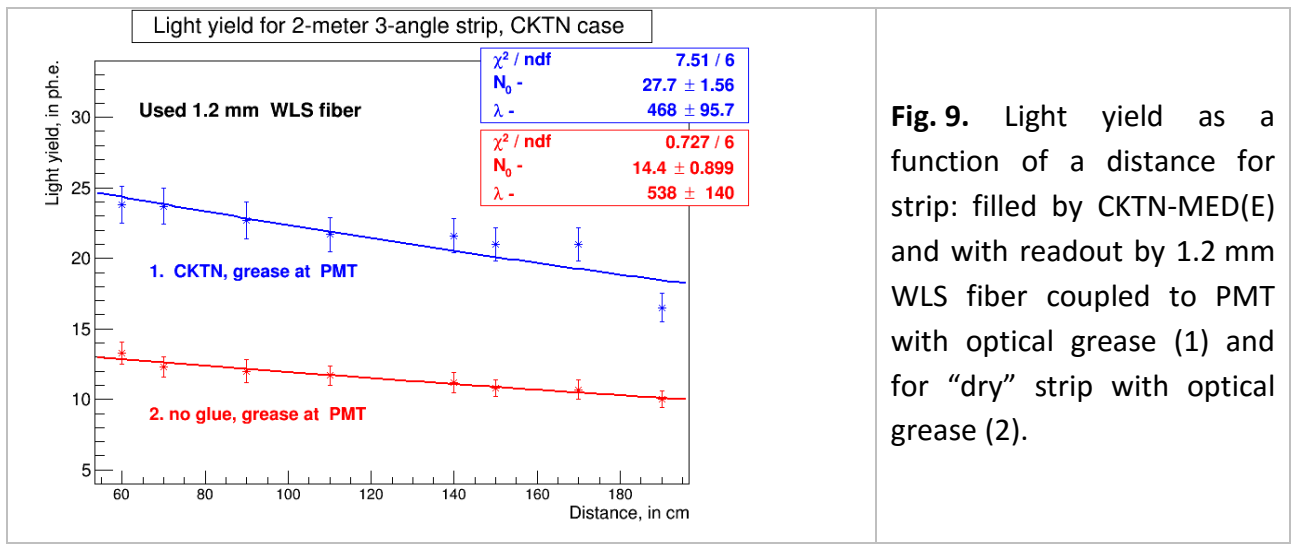

**Fig. 9.** Light yield as a function of a distance for strip: filled by CKTN-MED(E) and with readout by 1.2 mm WLS fiber coupled to PMT with optical grease (1) and for "dry" strip with optical grease (2).

We tested the dependence of the light yield on various diameters of WLS fiber and different reflectors at fiber far end (from PMT), etc. The study was done for strips with CKTN-MED(E) resin as a filler and with WLS fiber 1.0 mm or 1.2 mm diameter and results of these measurements are shown on Fig. 10 (curve 1 and 2 respectively). As one can see, light yield in "1.2 mm fiber" case is a factor 1.6-1.7 higher than that in "1.0 mm fiber" case. Similar result for the ratio of the light yield was obtained for Bicron WLS fibers with the 1.0 and 1.2 mm diameters [24].

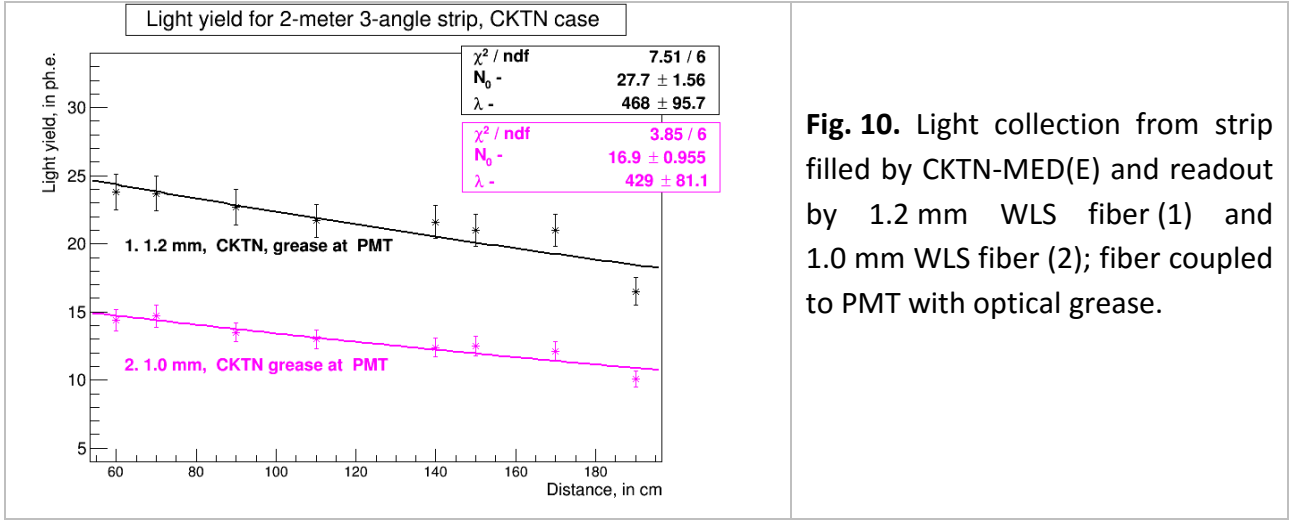

**Fig. 10.** Light collection from strip filled by CKTN-MED(E) and readout by 1.2 mm WLS fiber (1) and 1.0 mm WLS fiber (2); fiber coupled to PMT with optical grease.

Technical attenuation length (TAL) was found about 5 m for case with reflector on the far end and no obvious changes by using filler in the same other conditions. Another experiment was done when WLS fiber at the far end of the strip was covered by black paper in order to evaluate refractive light of this end. This study showed the significant decreasing of light collection with distances from PMT and its uniformity came too worse (curve 3 on Fig. 8). TAL is decreased more than 2 times (Fig. 8). This result points out to the importance of using high reflective material on the far end of WLS fiber.

But, in some cases, when time resolution of scintillation counter has a crucial importance, an application of blackened end of a fiber is possible to exclude a contribution of the reflected light. It improves time resolution but worsens the light yield from the strip. Filling the optical resin into the hole with WLS fiber should be one of the simple solutions to compensate of the light yield losses.

One can conclude that TAL for optical system "strip + WLS fiber" is mostly determined by WLS fiber with the mirror on the far end and badly decreased by a factor about 2 in case of the far end of the fiber was darkened. At the time, TAL for "1.2 mm fiber" case is about the same that the "1.0 mm fiber".

We also looked for the behavior of the light yield by time for the strip with 1.0mm WLS fiber and filled by CKTN-MED(E). No obvious differences in a light collection for this strip were observed over 6-month period: 14.0±0.7 and 13.9±0.7 photoelectrons at the middle part of strip (Table 3).

## 6. Conclusions

The light yield for 2 m long triangle shape scintillation extruded strips with WLS fiber filled by optical resins CKTN-MED(E) and BC-600 was studied on cosmic muons. These strips with 2.6 mm central co-extruded hole were made of polystyrene with dopants of 2% PTP and 0.03% POPOP; readout for the light of this strip was performed by WLS fiber inserted into the hole and coupled with PMT at one end.

We developed special technique to fill the co-extruded hole with WLS fiber inserted in by optical resins BC-600 or CKTN-MED (E).

Filling of the strip hole by the optical resin (CKTN-MED(E) or BC-600) gives the great increment of light yield by 1.6-1.9 times against to that of "dry" strip. Both optical resins showed almost similar results.

The comparison of the light yield for the strips with WLS fibers with 1.0 and 1.2 mm diameter was done as well. Increment of light yield using the WLS fiber with 1.2 mm diameter is almost 1.6 times more against to that of 1.0 mm diameter in the same other conditions.

Filling an optical resin into the hole with WLS fiber in it should be one of the simple solutions to compensate of the light yield losses when the blackened far end of a fiber is required to improve the time resolution for the strip counter.